\documentclass[11pt]{article}

\usepackage[T1]{fontenc}
\usepackage[utf8]{inputenc}
\usepackage{lmodern}
\usepackage{textcomp}
\usepackage{amsmath,amssymb}
\usepackage[margin=1in]{geometry}
\usepackage{graphicx}
\usepackage{microtype}
\usepackage[hidelinks]{hyperref}
\usepackage{doi}

\graphicspath{{figures/}}

\title{Finite State Machine-Based Traffic Orchestration for Multi-Engine Analysis Platforms}
\author{Ankit Kumar \and Thirumoorthi Thangamani}
\date{}

\begin{document}

\maketitle

\begin{abstract}
Modern application-security platforms route live traffic from a single
proxy into several independent analysis engines --- behavioral anomaly
detection, authentication analysis, resource discovery, access-control
inspection, among others. Each engine's appetite for traffic changes on
its own schedule, yet the proxy can honor only one forwarding
instruction per resource. We present a traffic-orchestration method that
models the \emph{combined} needs of all engines as a single composite
finite state machine (FSM) whose state space is the Cartesian product of
per-engine states. Every reachable composite state pre-maps to one
forwarding configuration, so runtime reduces to a lookup rather than a
negotiation. We add two refinements: time-based expiry treated as a
first-class FSM transition enabling declarative traffic decay, and
per-state multi-tier outputs that let one machine serve every service
level. The design is fully declarative, engine-agnostic, and scales to N
engines without re-architecture.
\end{abstract}

\section{Introduction}\label{sec:introduction}

Contemporary security platforms rarely rely on a single detector~\cite{denning1987,paxson1999}.
Instead they run a \emph{portfolio} of specialized analysis engines
behind a common ingress proxy or gateway. A behavioral engine learns
per-endpoint baselines and flags deviations~\cite{chandola2009}; an authentication engine
reasons about credentials and session anomalies; a discovery engine
catalogs newly observed APIs; an access-control engine checks
authorization semantics such as broken object-level authorization
(BOLA)~\cite{owaspbola2023}. These are representative, including but not limited to the
engine types named here --- a deployment may host additional engines,
and the method below is agnostic to how many there are. Each engine is
an independent consumer of the same underlying request stream.

The difficulty is that these engines do not want the \emph{same}
traffic, and they do not want it at the \emph{same time}. Traffic demand
is a per-engine, per-resource, and time-varying quantity. A behavioral
engine that has just flagged an endpoint as suspicious wants every
request for a while; the authentication engine may have already
concluded its analysis and want nothing; the discovery engine wants a
brief burst to seed its catalog and then falls silent. The proxy,
however, is a single control point: for any given resource it can
forward, sample, or drop~\cite{duffield2004,rfc5475,everflow2015} --- but it must commit to \emph{one} policy.

This paper addresses the coordination problem that sits between ``many
engines with divergent, drifting demands'' and ``one proxy that needs a
single, coherent answer per resource.'' Rather than letting engines
negotiate at runtime --- which produces race conditions, duplicated
taps, and incoherent policy --- we lift the entire coordination problem
into a \textbf{declarative composite finite state machine} that is
evaluated once per event and produces exactly one forwarding directive.

\section{Problem Statement}\label{sec:problem}

\begin{figure}[htbp]
\centering
\includegraphics[width=0.55\linewidth]{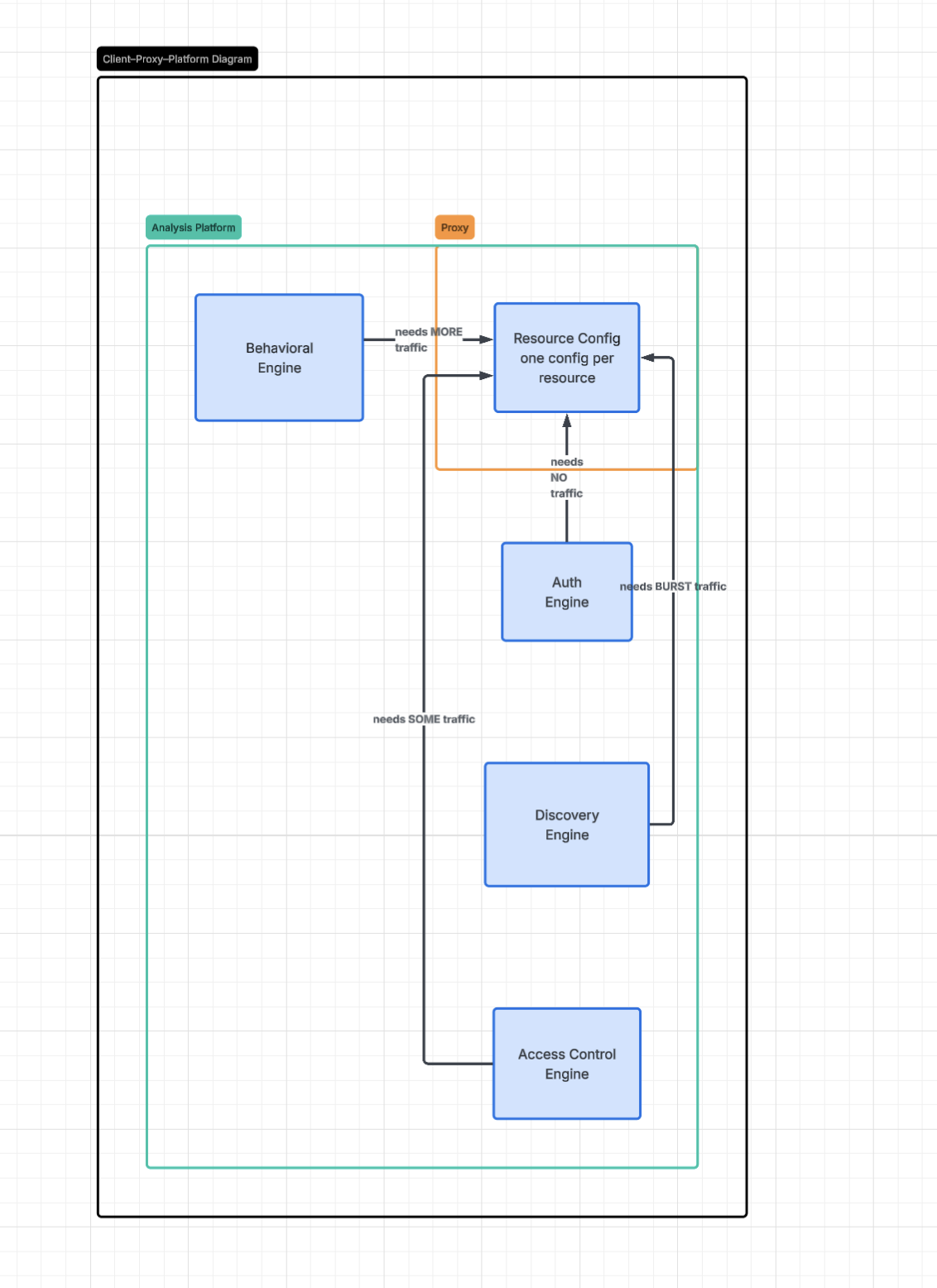}
\caption{\textbf{The core conflict.} Four engines express four
different, simultaneous demands against the \emph{same} per-resource
configuration: the behavioral engine ``needs MORE traffic,'' the
authentication engine ``needs NO traffic,'' the discovery engine ``needs
SOME traffic,'' and the access-control engine ``needs BURST traffic.''
The proxy sits at a single choke point and can publish only one config
per resource. The arrows converging on the ``Resource Config --- one
config per resource'' node make the contention explicit: multiple
writers, one slot.}
\label{fig:conflict}
\end{figure}

Consider the deployment in \textbf{Figure~\ref{fig:conflict}}. Traffic enters through a
proxy, which can mirror or forward it to an analysis platform hosting
four engines. Each engine annotates the shared \emph{resource
configuration} with what it currently needs.

The demands illustrated in Figure~\ref{fig:conflict} are not merely different in degree;
they are different in \emph{kind} (all, none, some, burst) and they
mutate independently as each engine advances through its own lifecycle.
This yields three hard requirements:

\begin{itemize}
\item
  \textbf{Single-writer coherence.} The proxy must receive exactly one
  unambiguous forwarding config per resource at any instant.
\item
  \textbf{Independence.} An event in one engine must not force other
  engines to re-announce their needs, and must not silently override
  them.
\item
  \textbf{Temporal drift.} Demand changes over time, sometimes on a
  fixed schedule (e.g., ``elevate for seven days, then step down''), and
  the system must handle that decay without any engine babysitting a
  timer.
\end{itemize}

Naive strategies each fail one of these. \emph{Forward everything}
satisfies coherence but wastes bandwidth and compute on engines that
need nothing~\cite{sekar2012,sherry2012}. \emph{Let each engine drive its own tap} satisfies
independence but destroys coherence, producing duplicated copies and
conflicting instructions. \emph{Highest-priority-wins} restores a single
answer but starves lower-priority engines that legitimately still need
\emph{some} traffic. What is missing is a principled way to
\textbf{compose} heterogeneous demands into one directive.

\section{Proposed Solution}\label{sec:solution}

\subsection{The composite state abstraction}\label{sec:composite}

We treat the union of engine lifecycles as a single machine. Each engine
contributes a \emph{dimension} --- a small set of discrete states
describing its current appetite. A resource's overall condition is
therefore a tuple, one coordinate per engine~\cite{harel1987,uml251}:

\begin{verbatim}
Behavioral :  BASELINE | SUSPICIOUS | THRESHOLD_DONE | POLICY_ACTIVE
Extended   :  DISABLED | ENABLED
Auth       :  STANDARD | NO_AUTH | ML_IN_PROGRESS | ML_COMPLETE
Access     :  IN_PROGRESS | DONE
\end{verbatim}

The reachable \textbf{composite state} is the Cartesian product of these
dimensions, e.g.
\texttt{(SUSPICIOUS,\ ENABLED,\ ML\_IN\_PROGRESS,\ IN\_PROGRESS)}. The
central design decision is that \emph{every} valid composite state is
pre-annotated, at definition time, with the exact forwarding
configuration the proxy should apply --- the tag set, the priority, and
flags such as bypass-suppression. The composite state is thus a
\textbf{traffic multiplexer}: it folds four independent opinions into
one deterministic output.

\begin{figure}[htbp]
\centering
\includegraphics[width=\linewidth]{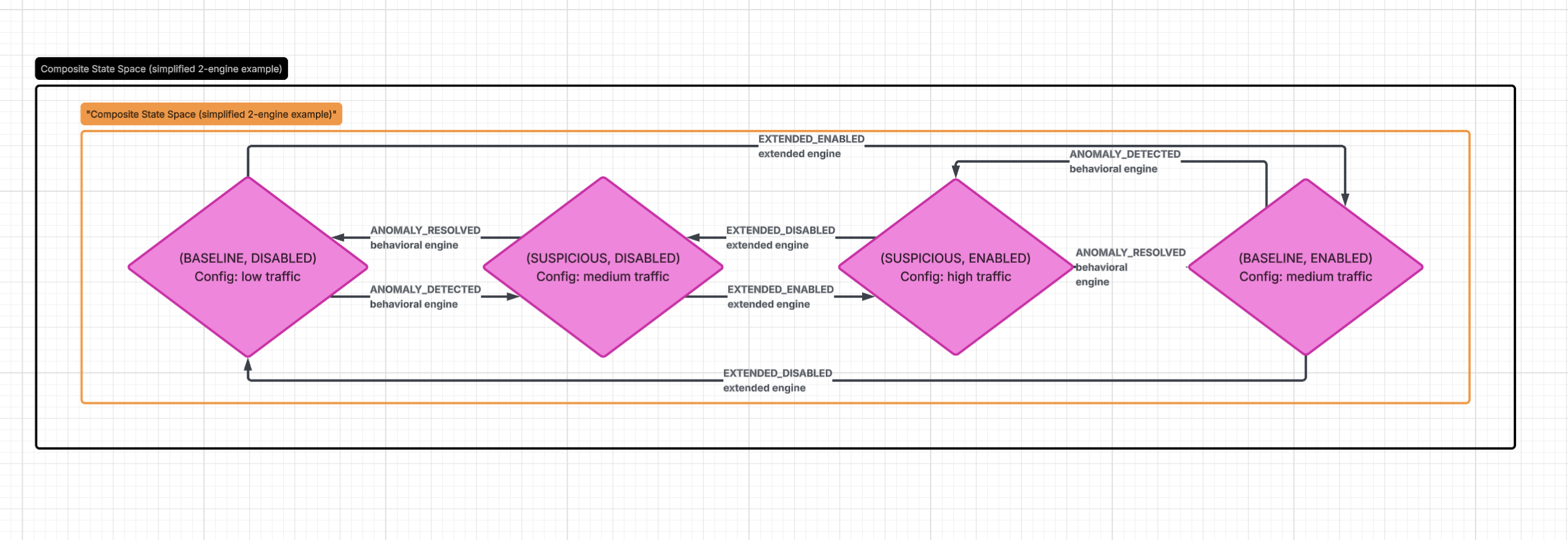}
\caption{\textbf{Composite state space and per-dimension transitions.}
Four composite states --- \texttt{(BASELINE,\ DISABLED)} $\rightarrow$ low traffic,
\texttt{(SUSPICIOUS,\ DISABLED)} $\rightarrow$ medium,
\texttt{(SUSPICIOUS,\ ENABLED)} $\rightarrow$ high, \texttt{(BASELINE,\ ENABLED)} $\rightarrow$
medium --- each carry a pre-computed config. Critically, every labeled
edge changes \textbf{only one coordinate}:
\texttt{ANOMALY\_DETECTED}/\texttt{ANOMALY\_RESOLVED} move the
behavioral dimension while leaving the extended dimension untouched, and
\texttt{EXTENDED\_ENABLED}/\texttt{EXTENDED\_DISABLED} do the reverse.
This orthogonality is what preserves engine independence: an event from
one engine can never clobber another engine's contribution, because it
is structurally incapable of touching the other coordinate. The full
production machine is the same picture in higher dimensions.}
\label{fig:statespace}
\end{figure}

\textbf{Figure~\ref{fig:statespace}} shows a simplified two-engine slice of this state
space (behavioral $\times$ extended) so the mechanics are visible without
combinatorial clutter.

\subsection{Runtime: event in, config out}\label{sec:runtime}

\begin{figure}[htbp]
\centering
\includegraphics[width=\linewidth]{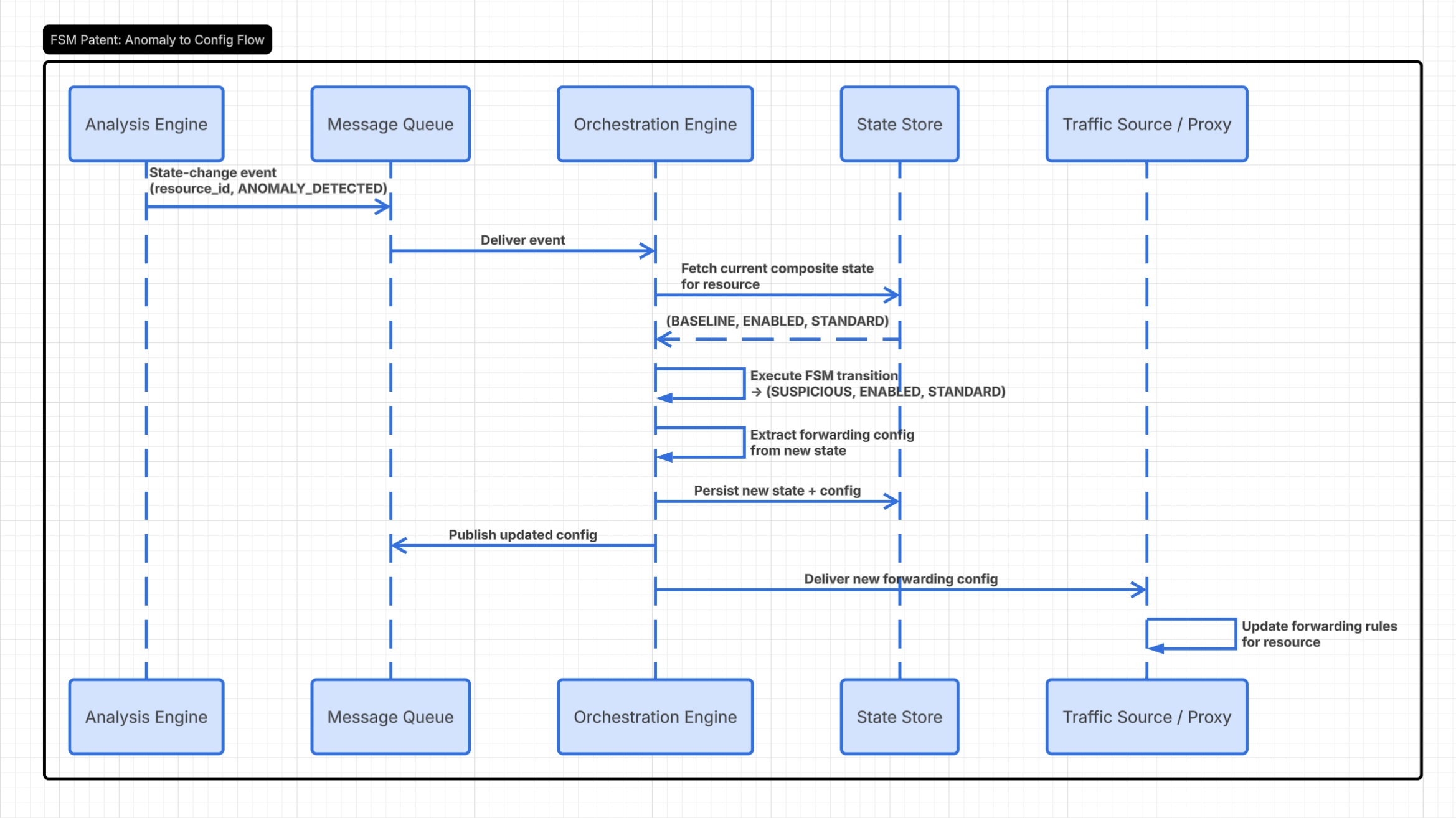}
\caption{\textbf{End-to-end event flow.} An analysis engine publishes a
state-change event (\texttt{\{resource\_id,\ ANOMALY\_DETECTED\}}) onto
a message queue. The orchestration engine consumes it, fetches the
current composite state from the state store (e.g.
\texttt{(BASELINE,\ DISABLED,\ STANDARD)}), executes the FSM transition
to the new state (\texttt{(SUSPICIOUS,\ ENABLED,\ STANDARD)}), extracts
the forwarding config from that state, persists the new state, and
delivers the resolved config to the proxy, which updates its forwarding
rules for the resource. No engine talks to another engine; all
coordination is mediated by the machine. The runtime performs a lookup,
not a computation --- the intelligence was compiled into the state
definition ahead of time.}
\label{fig:sequence}
\end{figure}

At runtime the system is a thin, generic executor. When any engine's
need changes it emits an event (\texttt{ANOMALY\_DETECTED},
\texttt{AUTH\_COMPLETE}, \texttt{DISCOVERY\_DONE}, \ldots) tagged with a
resource identifier. The orchestrator loads the resource's current
composite state, applies the single-dimension transition for that event,
resolves the new state's pre-mapped config, and pushes it to the proxy.
\textbf{Figure~\ref{fig:sequence}} traces this end-to-end.

\subsection{Expiry as a first-class transition}\label{sec:expiry}

\begin{figure}[htbp]
\centering
\includegraphics[width=0.75\linewidth]{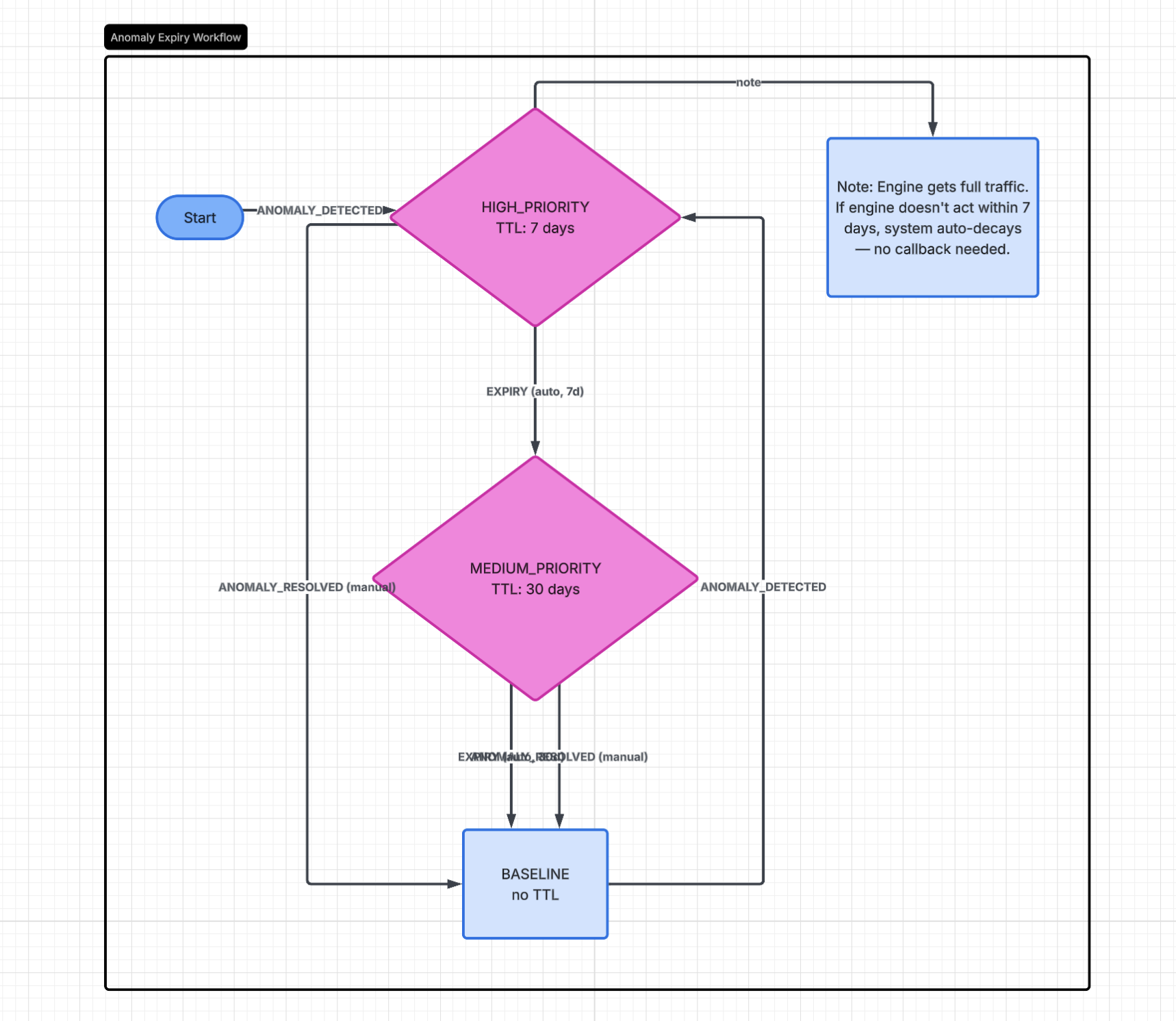}
\caption{\textbf{Expiry as declarative decay.} An
\texttt{ANOMALY\_DETECTED} event promotes the resource to
\texttt{HIGH\_PRIORITY} (full traffic, 7-day TTL). If the engine takes
no further action, the TTL fires an automatic \texttt{EXPIRY} transition
to \texttt{MEDIUM\_PRIORITY} (30-day TTL) and eventually to
\texttt{BASELINE} (indefinite). Manual signals such as
\texttt{ANOMALY\_RESOLVED} can short-circuit the curve at any point. The
key property --- noted in the diagram --- is that the engine gets full
traffic but \emph{need not call back}: the decay path lives in the FSM
definition, not in engine code. Expiry is a transition, not a deletion,
so the target state (and its config) is always well-defined.}
\label{fig:expiry}
\end{figure}

Some demands are inherently bounded in time: ``give this endpoint
elevated forwarding for seven days while I investigate, then ramp down
on your own.'' A conventional system implements this as a timeout that
\emph{deletes} state --- which is lossy and abrupt. We instead make
\textbf{expiry a normal FSM event}. Any state may declare a TTL~\cite{varghese1987}; when it
elapses the system fires an \texttt{EXPIRY} event that triggers a
\emph{defined} transition to a specific successor state. This lets a
whole decay curve be encoded declaratively:

\begin{verbatim}
HIGH_PRIORITY   (TTL: 7 days)
  --EXPIRY-->  MEDIUM_PRIORITY (TTL: 30 days)
    --EXPIRY-->  BASELINE (no TTL, indefinite)
\end{verbatim}

\textbf{Figure~\ref{fig:expiry}} renders this as a workflow. No engine manages a timer
or sends follow-up events; the machine ramps traffic down automatically.
In multi-instance deployments a distributed lock guarantees each expiry
fires exactly once~\cite{burrows2006} --- no duplicate transitions.

\subsection{One FSM, many tiers}\label{sec:tiers}

Customers subscribe at different service levels --- full analysis versus
basic monitoring. Rather than cloning the machine per tier, each
composite state carries \textbf{multiple named forwarding configs}, one
per tier:

{\small
\begin{verbatim}
State: (SUSPICIOUS, ENABLED, STANDARD, DISABLED)

  Full Analysis Tier:
    Tags: [DISCOVER, DETECT_BEHAVIORAL, DETECT_AUTH, DETECT_ACCESS, DETECT_EXTENDED]
    Priority: HIGH
    Bypass Suppression: true

  Basic Monitoring Tier:
    Tags: [DISCOVER, DETECT_BEHAVIORAL]
    Priority: NORMAL
    Bypass Suppression: false
\end{verbatim}
}

\begin{figure}[htbp]
\centering
\includegraphics[width=\linewidth]{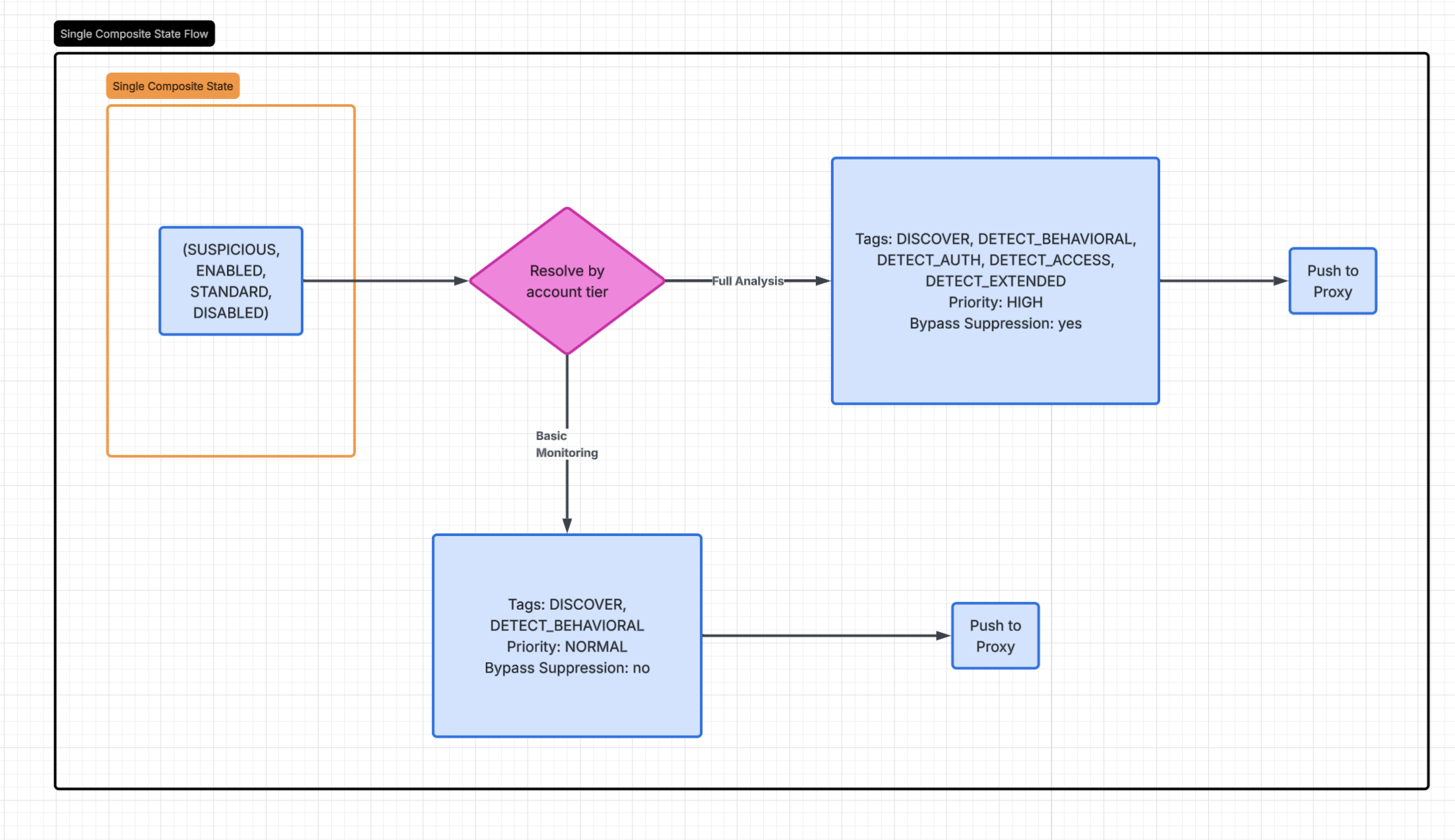}
\caption{\textbf{Tier resolution from a single composite state.} One
state, \texttt{(SUSPICIOUS,\ ENABLED,\ STANDARD,\ DISABLED)}, feeds a
``resolve by account tier'' decision. The Full Analysis branch emits the
rich tag set at HIGH priority with suppression bypassed; the Basic
Monitoring branch emits a reduced tag set at NORMAL priority. Both paths
end at ``push to proxy.'' Because the machine's structure is shared,
adding or retuning a tier is a change to per-state annotations, not a
new state machine.}
\label{fig:tiers}
\end{figure}

At resolution time the orchestrator reads the resource owner's tier and
selects the matching config. The transition logic --- states, events,
edges --- is entirely tier-agnostic; only the \emph{output} differs.
\textbf{Figure~\ref{fig:tiers}} shows this selection step.

\subsection{Declarative definition}\label{sec:declarative}

The entire machine --- states, transitions, per-state configs, TTLs, and
tier overrides --- is expressed in a structured configuration file
(e.g., YAML), not in code~\cite{scxml2015,asl,burns2016}. The runtime is a generic engine: load
definition, receive event, transition, extract config, publish. It knows
nothing about security or traffic semantics. Adding a new engine
dimension means adding states and edges to the config; adjusting a decay
curve means editing TTLs; tier differences are visible in one place.
Domain intelligence lives in the definition, not the executor.

\section{Advantages}\label{sec:advantages}

\textbf{Deterministic single-writer coherence.} Because each composite
state maps to exactly one config, the proxy always receives an
unambiguous directive --- the multi-writer contention of Figure~\ref{fig:conflict} is
resolved by construction, not by locking or last-writer-wins heuristics.

\textbf{Engine independence with no runtime negotiation.}
Single-dimension transitions (Figure~\ref{fig:statespace}) mean an event from one engine
cannot overwrite another's contribution. Engines never coordinate
directly, eliminating negotiation latency, redundant taps, and
coordination races.

\begin{figure}[htbp]
\centering
\includegraphics[width=\linewidth]{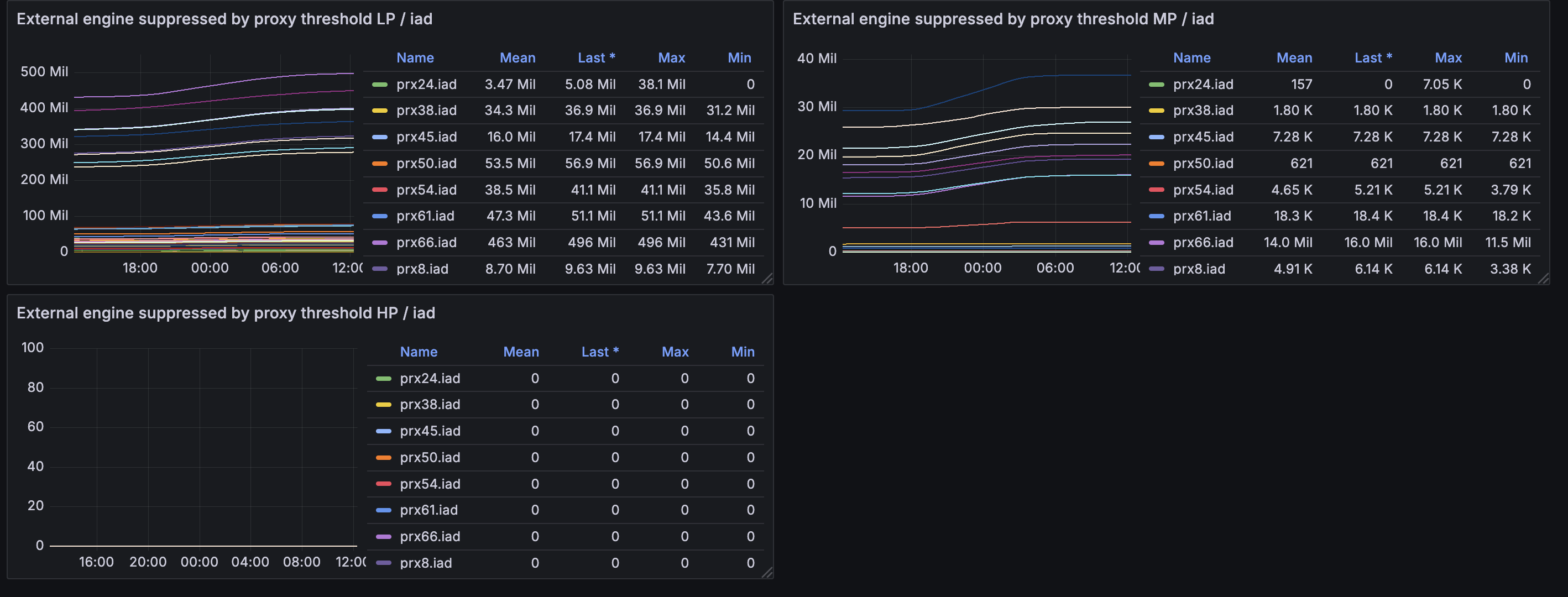}
\caption{\textbf{Priority-driven suppression, measured in production.}
Three Grafana panels plot the volume of external-engine traffic
\emph{suppressed by the proxy} at each priority threshold --- Low (LP),
Medium (MP), and High (HP) --- broken out per proxy instance
(\texttt{prx24.iad}, \texttt{prx38.iad}, \ldots). At the \textbf{LP}
threshold the proxy suppresses hundreds of millions of requests: traffic
that no engine currently needs is dropped at the edge rather than
forwarded and inspected. At \textbf{MP} suppression falls by roughly
four orders of magnitude (thousands to low millions), and at \textbf{HP}
it is effectively \textbf{zero} --- high-priority traffic, which some
engine has explicitly escalated to, is always forwarded. This is the
FSM's forwarding config operating as designed: the priority carried by
each composite state (Section~\ref{sec:tiers}) becomes the proxy's suppression
threshold, so compute is spent only where an engine has expressed
demand. The gap between the LP and HP panels is the concrete, quantified
cost saving of the composite-state approach.}
\label{fig:suppression}
\end{figure}

\textbf{Precise forwarding, lower cost.} Only traffic that some engine
actually needs is forwarded, at the priority it needs. This avoids the
``forward everything'' tax on bandwidth and compute and reduces
downstream noise (fewer false positives from engines inspecting traffic
they don't want). \textbf{Figure~\ref{fig:suppression}} shows this effect in production: the
priority a composite state resolves to directly governs how much traffic
the proxy suppresses versus forwards.

\textbf{Adaptive, self-escalating defense.} Expiry-as-transition (Figure~\ref{fig:expiry}) gives declarative traffic decay: the platform elevates protection on
detection and ramps down automatically, with no engine-side timers and
exactly-once semantics under a distributed lock.

\textbf{One machine serves all service levels.} Multi-tier outputs
(Figure~\ref{fig:tiers}) collapse what would be N parallel state machines into one,
cutting maintenance surface and guaranteeing that all tiers share
identical, audited transition logic.

\textbf{Scales to N engines as configuration, not code.} New
capabilities are new dimensions and edges in the declarative definition.
The generic runtime is untouched, so the system grows without
re-architecture --- new capability equals config, not code.

\section{Conclusion}\label{sec:conclusion}

The friction in multi-engine analysis platforms is not detection quality
but \emph{coordination}: many specialized engines, each with an
independent and time-varying appetite for traffic, all funneling into a
single proxy that can honor only one directive per resource. We resolved
this by modeling the engines' combined demand as a composite finite
state machine whose state space is the Cartesian product of per-engine
states, with every reachable state pre-mapped to a single forwarding
configuration. Three properties make the approach distinctive: composite
state as a traffic \emph{multiplexer} that folds many opinions into one
deterministic output; expiry treated as a first-class transition,
enabling declarative, cascading traffic decay without engine-managed
timers; and per-state multi-tier annotations that let a single machine
serve every service level. Because the machine is fully declarative and
the runtime is domain-agnostic, the system scales to additional engines
and tiers through configuration alone. The net effect: independent,
asynchronous, time-varying engine demands yield a single, coherent,
self-managing forwarding directive per resource --- with no engine ever
needing to know that any other engine exists.

\end{document}